# Large-Scale Vandalism Detection with Linear Classifiers

## The Conkerberry Vandalism Detector at WSDM Cup 2017

Alexey Grigorev
Searchmetrics GmbH
contact@alexeygrigorev.com

## ABSTRACT

Nowadays many artificial intelligence systems rely on knowledge bases for enriching the information they process. Such Knowledge Bases are usually difficult to obtain and therefore they are crowd-sourced: they are available for everyone on the internet to suggest edits and add new information. Unfortunately, they are sometimes targeted by vandals who put inaccurate or offensive information there. This is especially bad for the systems that use these Knowledge Bases: for them it is important to use reliable information to make correct inferences.

One of such knowledge bases is Wikidata, and to fight vandals the organizers of WSDM Cup 2017 challenged participants to build a model for detecting mistrustful edits. In this paper we present the second place solution to the cup: we show that it is possible to achieve competitive performance with simple linear classification. With our approach we can achieve AU ROC of 0.938 on the test data. Additionally, compared to other approaches, ours is significantly faster. The solution is made available on GitHub[1].

## 1. INTRODUCTION

Knowledge bases provide many information processing systems with important data: these systems can use the knowledge bases to enrich the information they deal with and based on that understand the information better.

There are popular open Knowledge Bases such as Wikidata: it is community driven and everybody can add anything they want. Unfortunately, this base is sometimes targeted by vandals – the people who put inaccurate or offensive information there.

Such edits are quite harmful for the systems that use Wikidata: they rely on its content and may cause the algorithms to output unexpected results or come to inaccurate conclusions. To remove such edits, Wikidata's moderators have to constantly monitor incoming edits and roll back the vandalic ones.

One of the challenges offered at WSDM Cup 2017 was the Vandalism Detection challenge [6]: the task was to predict whether a Wikidata revision should be rolled back or not. For that the historical data with 72 million previous revisions was provided, and all the revisions which needed to be rolled back were marked [4].

In this paper we present the approach which took the second place at this task. For solving this challenge we used the hashing trick and linear SVM models and we show that for such large scale classification task simple linear models are very competitive, and, what is more, are quite fast.

The solutions to the challenge were evaluated on the TIRA platform [11], which allowed to test the solution close to the real-world conditions, where the revisions are streamed to the model as they come in, and the model makes predictions in real time and sends the results back. Upon execution the platform reported metrics such as accuracy, precision, recall, $F_1$ score, PR AUC, AU ROC and execution time. The final standings of the contestants was based on the performance of the models measured by AU ROC.

Prior to the competition there has been some research to automatic vandalism detection in knowledge bases: Heindorf and others have extracted the present dataset from Wikidata with reverted revisions and proposed a baseline solution [5].

## 2. DATASET DESCRIPTION

There are several datasets provided for the competition: the Wiki-Data XML dump files, the csv files with meta information about users and the label information (whether a revision was reverted or not).

The 22 xml dump files with more than 72 million revisions are distributed in form of 7z archives, which took 24.8 GB of disk space in the compressed form, or more than 400 GB uncompressed.

A typical revision looks like this:

```
<page>
  <title>Q5704066</title>
  <ns>0</ns>
  <id>5468191</id>
  <revision>
    <id>185142906</id>
    <parentid>185051367</parentid>
    <timestamp>2015-01-01</timestamp>
    <contributor>
      <username>{{USERNAME}}</username>
      <id>52267</id>
    </contributor>
    <comment>{{COMMENT}}</comment>
    <model>wikibase-item</model>
    <format>application/json</format>
    <text xml:space="preserve">
      {{PAGE_JSON}}
    </text>
  </revision>
</page>
```

In this snippet we see that a revision is contained in a page tag which has the title and the id, and then the revision itself has the id, the timestamp when the edit was made, the information about the user who suggested the edit (or IP in case the user was anonymous), the comment text and the entire revision content in JSON (in the snippet above the comment text and the JSON are removed for brevity).

---

[1]https://github.com/wsdm-cup-2017/conkerberry and https://github.com/alexeygrigorev/wsdmcup17-vandalism-detection

Often the users are not logged in to Wikidata, and in this case the meta file contains the following information about them:

- continent code;
- country code;
- region code;
- county;
- city;
- time zone.

Also, the meta file contains the information about revision tags.

Finally, there is a separate dataset with information about which revisions were reverted. According to this file, only 187 thousand revisions were reverted, or only 0.25% of total revisions. This poses an additional challenge: the amount of positive examples (rollbacks) are very small and hence the dataset is extremely skewed.

Next, we will describe our solution in more details.

## 3. APPROACH

In this section we present our approach to the challenge in details. First, we describe the hardware and software used for the solution, then we show the validation scheme used in our experiments, and finally we talk about the features which we extracted from the challenge dataset as well as the models built on these features.

### 3.1 Environment

The experiments were performed on a Linux Ubuntu server with 32GB RAM and 8 Cores.

We used Python 2.7 and the PyData stack for our development:

- `numpy` 1.11.2 for numerical operations [12];
- `scipy` 0.18.1 for storing sparse data matrices [7];
- `pandas` 0.17.1 for tabular data manipulation [9];
- `scikit-learn` 0.18.1 for data preprocessing and machine learning [10].

We used Anaconda – a distribution of Python with many scientific libraries pre-installed[2], including numpy, scipy, pandas and scikit-learn.

Additionally, we used the following libraries:

- `lxml` 3.7.0 for processing the xml dump files with Wikidata revisions [1];
- `feather` 0.3.1 for storing the processed data[3];
- `twisted` 16.5.0 for interacting with TIRA via sockets [8].

Note that the TIRA platform imposes the following limitation on the solution: it provided only one CPU and the solution should be able to run with 4 Gb of RAM or less.

---

### 3.2 Validation

The provided dataset was split into three parts by the organizers: the training set, the validation set and the testing set. The testing part was not released to the contest participants during the challenge, and it was only used for the final evaluation via the TIRA platform.

The split of the dataset is time-based:

- **training**: from 2012-10-29 to 2016-02-29 (65 mln revisions);
- **validation**: from 2016-03-01 to 2016-04-30 (7.2 mln revisions);
- **testing**: from 2016-05-01 onwards (10.4 mln revisions).

Because of the size of the dataset it was not possible to use the entire training data on our hardware. To overcome this problem we used only the data from 2015 and 2016 years and did not take into account the revisions before 2015.

Since we did not have access to the actual testing set, in our experiments we used the validation set as the testing set, and split the provided training set manually into training and validation subparts. Thus, for our models we used the following split:

- **training**: from 2015-01-01 to 2015-31-31 (27.9 mln revisions);
- **validation**: from 2016-01-01 to 2016-02-29 (9.3 mln revisions);
- **testing**: 2016-03-01 to 2016-04-30 (7.2 mln revisions).

In this section and the next ones, "testing" will refer to the dataset we used in the local experiments – i.e. with revisions from 2016-03-01 to 2016-04-30, which corresponds to the "validation" dataset of the cup.

The training part was used for training the model, and the validation part was used for tuning the hyperparameters of the models. The testing dataset was used only occasionally to verify that the approaches do not overfit.

During the split we noticed that the amount of positive examples is even smaller for the parts of data we selected: there are only 0.18%, 0.10% and 0.14% rollbacks for training, validation and testing subsets respectively.

### 3.3 Features

In [5] Heindorf and others describe a number of features for solving the vandalism detection task. However, in our experiments we did not base the feature engineering process on any prior work.

For this challenge we propose to use three categories of features:

- **Page-based features**: the information about the page such as Page Title;
- **User-based features**: the information about the user such as user name or meta features if the user is not logged in;
- **Comment-based features**: the information from the comment of the revision.

Content-based features based on the JSON information were not considered for this task as it was too complex to extract them efficiently in real time under the 4 Gb RAM constraint. This is why these features were not computed.

While the page-based features are rather simple – just the page's ID as a feature – other are more complex. Let us take a closer look at them.

**User-based Features**

For users who are logged in we use the username of the user. If the user is not logged in, we add a feature `anonymous=true` and take all the geographical information about the user such as continent, country, city and timezone.

For example, for an anonymous user from Leeds, UK we will generate the following features:

- `country_code=GB`
- `continent_code=EU`
- `time_zone=GMT`
- `region_code=EN`
- `city_name=LEEDS`
- `county_name=WEST_YORKSHIRE`

Additionally, we use extract "path" features from the IP of the anonymous user. For example, for an IP `90.219.230.105`[4] we would generate the following features: `90`, `90_219`, `90_219_230`, `90_219_230_105`.

Revision tags are also included to the set of User-based features, but we perform no additional preprocessing of them.

**Comment-based Features**

Another set of features is extracted from the revision comments. A typical Wikidata revision comment is system-generated (see fig. 1). As we see, the comment is quite structured: there is some technical description inside `/* */`. To extract features from the technical part we split this string by "`:`" and "`|`" characters and use the results as tokens.

Also there is the text that users add to the page, and it is located outside `/* */`. We can tokenize this text and use it as text features.

Finally, for some comments we have the information about the predicate of the relation that is added to the knowledge base (the "Property" link) and the object of the relation (the link after the predicate). These wiki-links can be extracted by looking inside all the double squared brackets and extracting the text from there.

Thus, we can generate three sets of features from the comment:

- **Structured comment**: the information inside `/* */`;
- **Links**: all the wiki-links inside the comment;
- **Unstructured comments**: the text tokens outside `/* */`.

Note that not all the comments follow this pattern, and in case a comment is different we just leave the corresponding feature empty.

### 3.4 Models

Next, we built a model for each feature separately. The basic algorithm for doing it is the following:

1. Take a feature and produce a set of One-Hot-Encoded vectors by tokenizing the feature strings with the `CountTokenizer` class from `scikit-learn`. The tokenization is performed on the word level and can be trained on the entire dataset. This step produces a sparse `scipy` matrix;

---
[4] the IP is not real and only used for demonstration purposes

**Table 1: The performance of models built on different features**

| Feature | $C$ | Time | AU ROC |
|---------|-----|------|--------|
| Title feature | 0.5 | 30 sec | 0.64 |
| User features | 0.1 | 10 min | 0.93 |
| Struct. comment | 0.1 | 25 min | 0.89 |
| Link features | 10 | 8 min | 0.72 |
| Unstruct. comment | 1 | 15 min | 0.83 |
| Final model | 0.5 | 45 min | 0.96 |

**Table 2: The performance of ensemble classifiers**

| Model | Validation | Test |
|-------|-----------|------|
| User Model | 0.932 | 0.951 |
| LogReg | 0.951 | 0.941 |
| XGBoost | 0.971 | 0.950 |
| Final model | 0.966 | 0.960 |

2. Fit a machine learning model to the matrix produced on the previous step;

3. Evaluate the performance of the model on the validation dataset.

With this we created the following five models:

- title model;
- user model;
- models based on comment features:
  - structured comments model;
  - links model;
  - unstructured comments model.

At this step we tried several algorithms and settled down on Linear SVM classifier in the primal form with $L_1$ regularization optimized via LIBLINEAR [3] (the `LinearSVC` class in scikit-learn).

This model has one parameter $C$ which specifies the amount of $L_1$ regularization we impose on the model's weights.

The parameter $C$ was tuned for each model separately with the gridsearch strategy: we tried a wide range of parameter candidates starting from $10^{-7}$ up to 10. The performance (by AU ROC metric) of each produced model was evaluated on the validation set and the best performing model was used. For the results refer to Table 1.

**Ensembling**

The initial idea was to use the output of the above models as meta-features and train another model on top of this. This approach is called stacking [14] and is very popular on Data Science challenge platforms like Kaggle[5]: it allows to get a strong second-level model by combining other models and significantly improve the score.

We tried two second level models for this challenge:

- Gradient Boosted Trees as implemented in XGBoost [2];
- $L_2$-regularized Logistic Regression with LIBLINEAR optimizer [3].

However, stacking did not work well for this task: initially it appeared promising and was able to achieve 0.97 AU ROC with

---
[5] http://kaggle.com/

Example 1: /* wbsetdescription-add:1|es */ futbolista irlandes
Example 2: /* wbcreateclaim-create:1| */ [[Property:P31]]: [[Q5]], #autolist2

**Figure 1: Examples of comment data from Wikidata revisions.**

XGBost (as measured by 2-fold cross-validation with random shuffling).

However, when we compared the best single model and the ensemble on testing, it turned out that they comparable in performance (see table 2). Clearly, the XGBoost model overfits the training data and its behavior on unseen data is unpredictable, hence it is quite risky to use, and simpler models should be preferred.

One possible explanation why it did not perform well is the extreme skewness of the data: both models captured noisy fluctuations of the positive examples in the data and overfit on them.

**Online learning**

Because of the hardware limitations we had to give up a large amount of data and use only approximately 50% for training the model.

Typically this limitation can be overcome by using online learning methods: these methods do not need to access the entire dataset at once, but instead consider only a small batch at a time and gradually update the parameters as they process the data.

One of such approaches is to use Stochastic Gradient Descent base models, which are implemented in `SGDClassifier` class in scikit-learn. If we use the hinge loss for such a classifier, then it is equivalent to leaning the primal form of Linear SVM.

However, our experiments showed that the model behavior is very unstable during the training. Typically this is solved by decreasing the learning rate of the model, but in this case the model stopped learning anything.

We suspect that this also due to high skewness of the data: the positive examples are too rare to make any effects and there are too few of them in a batch.

**Upsampling and Resampling**

Since we previously observed that the positive examples are too rare we wanted to experiment with ways to overcome this and make the dataset more balanced.

There are many ways of doing this, but since our feature vectors are very sparse and binary, the are not available options. In our experiments we focused on the following two approaches:

- Upsampling the positive class: sampling the reverted edits with replacement;

- Downsampling the negative class: sampling with or without replacement the good edits.

However, experiments showed that none of these approaches was better than considering the entire dataset as is: the best performing model was able to achieve only AU ROC of 0.90.

**Feature Hashing**

For the final model we decided to combine all the features into one large feature string, tokenize it and train one single SVM on them.

If we do this, then our feature space becomes very large and it is quite expensive to store the mapping from the string feature representation to its index in a sparse matrix. We also suspected that it may cause problems when evaluating the model on the TIRA platform: when stored on disk there models occupied 2.5 Gb of space (both tokenizers and the weights learned by SVMs). Our concern was that 1.5 Gb might be not enough for efficient processing of incoming data.

| Team name | AU ROC | Running Time |
|-----------|--------|--------------|
| buffaloberry | 0.947 | 17:11:16 |
| **conkerberry** | 0.937 | 02:47:50 |
| loganberry | 0.919 | 104:47:30 |
| honeyberry | 0.904 | 26:37:29 |
| riberry | 0.894 | 189:16:03 |

**Table 3: Top 5 participants of the challenge.**

We can overcome the need of storing a dictionary by using the hashing trick [13]: instead of creating such a dictionary, we find the needed column index by computing the hash of the token.

With this approach the vectorizer uses virtually no memory, and the model parameters of the resulting SVM occupy only 300 Mb, making the memory footprint of our application very low.

This combined model achieved the AU ROC with $C = 0.5$ and scored 0.966 and 0.960 on the validation and testing set respectively. Training such a model took 45 minutes.

## 4. EVALUATION RESULTS

For the final evaluation we selected the combined SVM model and executed it against the withheld testing data on the TIRA platform.

With this model our team was able to achieve the second position in the final standings with AU ROC of 0.937 (see table 3: our team is in bold).

Because not all the information was used, and the classifier is quite simple, the model achieved 1% AUC less than the first place solution. But our approach was the fastest: on the training set it finishes in 3 hours, while the next fastest solution is approximately 6 times slower.

## 5. CONCLUSION

In this paper we showed that linear classifiers are very competitive models for large scale prediction problems like the Vandalism Prediction challenge, and it outperforms more complex models like ensembles and Gradient Boosted Trees.

The hashing trick allows to keep the memory footprint very low, and combining it with Linear SVM classifier enables to score incoming revisions very quickly. With a speedup of 6 times it was only worse by 1% to the first place solution.